\documentclass[11pt]{article}

\usepackage[utf8]{inputenc}
\usepackage[T1]{fontenc}
\usepackage{lmodern}
\usepackage[margin=1in]{geometry}
\usepackage{amsmath,amssymb}
\usepackage{graphicx}
\usepackage{booktabs}
\usepackage{array}
\usepackage{caption}
\usepackage{authblk}
\usepackage[hidelinks]{hyperref}
\usepackage{xurl}

\title{\textbf{Disentangling Answer Engine Optimization from Platform Growth:\\
A Log-Based Natural Experiment on ChatGPT Referral Traffic}}

\author[1]{Keisuke Watanabe\thanks{Equal contribution; co-corresponding author (\texttt{kei@glasp.co}).}}
\author[1]{Kazuki Nakayashiki\thanks{Equal contribution; co-corresponding author (\texttt{kazuki@glasp.co}).}}
\affil[1]{Glasp Inc.}
\date{\today}

\begin{document}
\maketitle

\begin{abstract}
\noindent
Large language model (LLM) ``answer engines'' such as ChatGPT now send measurable referral
traffic to the open web, and a practice analogous to search engine optimization, here called
\emph{Answer Engine Optimization} (AEO), has emerged. Public AEO success stories typically
quote large raw growth multiples, but raw referral growth is confounded by the rapid
platform-level growth of the answer engines themselves. We report a longitudinal field study
on a single high-traffic domain (\texttt{glasp.co}) whose corpus of hundreds of thousands of YouTube
question-and-answer pages received a defined bundle of AEO interventions in January 2026
(detailed in Section~\ref{sec:treatment}).
Because the interventions were concentrated on one subset of the site, the untreated remainder
of the same domain acts as a contemporaneous control that absorbs the platform tailwind.
Using first-party analytics and server logs rather than probabilistic third-party estimators,
we find: (1) raw growth is dominated by the platform tailwind: on monthly aggregates total
ChatGPT referrals grew $5.7\times$ while \emph{untreated} pages on the same domain grew $3.5\times$
over the same window;
(2) an interrupted time-series model on the weekly treated/control ratio estimates a
discrete, intervention-aligned level increase of $\times 1.82$ (95\% CI $1.31$--$2.54$,
HAC $p{=}0.001$), robust across engagement-filtered traffic ($\times 2.27$) and alternative
specifications; (3) however, a conservative placebo-in-time permutation test yields
$p{=}0.16$, so the effect is \emph{suggestive, not conclusive}, given a short and noisy
pre-period; and (4) Google organic clicks to treated pages did not fall beyond the ambient
site-wide trend and indexation was preserved, consistent with the SEO-protection rule.
The methodological message, separating treatment from platform tailwind with an on-domain
control, matters more than any single multiple, and implies that headline AEO multiples
substantially overstate causal effect.
\end{abstract}

\section{Introduction}
LLM-based answer engines increasingly mediate how users reach web content. When ChatGPT,
Perplexity, Gemini, or Copilot answer a question, they may cite and link sources, producing a
referral channel distinct from classic web search. Practitioners have begun optimizing content
for this channel, a practice we call Answer Engine Optimization (AEO), parallel to and partly
overlapping with search engine optimization (SEO) and the recently formalized notion of
Generative Engine Optimization~\cite{geo2024}.

Public AEO case studies typically report a single dramatic growth multiple. Such multiples are real
but difficult to interpret, because
the answer engines themselves grew explosively over the same window, and independent evidence
shows AI answer surfaces are simultaneously reshaping click behavior on the open
web~\cite{pew2025}. The central question is therefore not ``did traffic grow,'' but
\textbf{how much of the growth is attributable to the optimization versus to the platform
tailwind that lifts all sites.}

We answer this for a single domain using a natural experiment. The AEO interventions were
applied to one identifiable subset of the site (a corpus of YouTube question-and-answer pages),
while the rest of the same domain was left untreated. The untreated subset is exposed to the
identical platform tailwind, the same domain authority, and the same analytics and
bot-filtering regime, and therefore functions as a contemporaneous control.

\paragraph{Contributions.}
\begin{enumerate}
  \item A \textbf{log-and-analytics measurement methodology} for AEO that uses first-party
  server logs and web analytics rather than probabilistic third-party citation estimators.
  \item A \textbf{natural-experiment design} that nets out platform-level growth using an
  on-domain control, analyzed with an interrupted time series (ITS) and a placebo-in-time test.
  \item A \textbf{conservative, range-reported effect estimate}, which we argue is the honest
  unit of an AEO result, in contrast to confounded raw multiples.
  \item A \textbf{validated SEO-safety mechanism} showing that targeted rewriting need not
  damage existing organic search performance.
\end{enumerate}
We do not claim a randomized effect: the interventions were a product decision, not a
randomized trial, and we report the resulting threats to validity in full (Section~\ref{sec:threats}).

\section{Background and Related Work}
\paragraph{Generative and Answer Engine Optimization.}
Aggarwal et al.~\cite{geo2024} introduced Generative Engine Optimization and showed, in a
controlled synthetic benchmark, that content modifications (citations, statistics, authoritative
phrasing) can raise a source's visibility \emph{within generated answers}. Our work is
complementary in method: rather than a synthetic benchmark of generation visibility, we measure
realized downstream \emph{referral traffic} on a live domain over time.

\paragraph{Citation behavior of answer engines.}
Liu et al.~\cite{liu2023verifiability} audited citation recall and precision of early generative
search engines, establishing that what answer engines cite is measurable and uneven. Whether a
source is cited (their object of study) and whether users then click through (ours) are distinct
links in the same chain.

\paragraph{AI answer surfaces and the open web.}
Pew Research Center~\cite{pew2025} found that users click linked sources far less often when an AI
summary is present, motivating why click-through from answer engines, not merely citation, is the
quantity practitioners ultimately care about.

\paragraph{Methodology.}
Our identification borrows the segmented-regression ITS framework of Wagner et al.~\cite{wagner2002}
and the difference-in-differences (DiD) logic, while heeding the warning of Bertrand et
al.~\cite{bdm2004} that serial correlation inflates significance in such designs. We therefore use
HAC (Newey--West~\cite{neweywest1987}) standard errors and a placebo-in-time permutation test.

\section{Setting and Data}
\paragraph{Site and corpus.} \texttt{glasp.co} is a social web highlighter. The relevant corpus is
hundreds of thousands of YouTube question-and-answer pages under the path prefix \texttt{/youtube/}. These
pages present transformative question--answer summaries of public videos with creator attribution
and embedded video, not republished transcripts.

\paragraph{Treated vs.\ control.} The AEO interventions (Section~\ref{sec:treatment}) were applied
to the \texttt{/youtube/} corpus. All other paths on the same domain (long-form articles, discovery
pages, user profiles, reader) received none of the interventions and constitute the control group.
Both groups share the domain, the platform tailwind, the analytics configuration, and the
bot-filtering regime.

\paragraph{Data sources and definitions.} Outcomes are computed from the Google Analytics~4
Data API: ChatGPT referrals are sessions whose \texttt{sessionSource}
contains \texttt{chatgpt} (dominated by \texttt{chatgpt.com}; other AI engines together contribute
$<$1\% of AI-referred sessions in the study window and are excluded). The landing-page path splits
treated vs.\ control. We report both raw \emph{sessions} and GA4 \emph{engaged sessions} (the
latter is more robust to non-human referral noise). SEO outcomes (clicks, impressions) come from
Google Search Console (\texttt{sc-domain:glasp.co}) for pages containing \texttt{/youtube/}.

\paragraph{A note on data precision.} GA4 totals are not perfectly additive across query
granularities (thresholding/approximation). Rebuilding monthly totals from a daily pull and
comparing against an independent monthly query, we observe a maximum discrepancy of $3.3\%$.
We therefore use the daily-derived series as the single canonical source for the time-series
analysis and treat differences below this level as noise. All May 2026 figures are through
2026-05-28 (partial month).

\paragraph{Confidentiality and scale.} To protect commercially sensitive volumes, this paper reports
all quantities in \emph{relative} form: growth multiples, group ratios, and series indexed to
January 2026 ($=100$). Absolute session, page, and impression counts are withheld; the accompanying
public release contains indexed aggregates and analysis code, not raw counts. To convey statistical
power without exact figures: total monthly ChatGPT referral sessions reached the hundreds of
thousands by May 2026, the treated corpus comprises hundreds of thousands of pages, and the control
group sustained on the order of thousands to tens of thousands of sessions per week throughout the
window, comfortably above the level at which the weekly treated/control ratio is stable. Withholding
absolutes changes no result, because every estimate (the level break, the platform-tailwind factor,
the difference-in-differences, the SEO comparison) is scale-invariant.

\section{Interventions: the AEO Treatment}\label{sec:treatment}
The treatment is a bundle applied to the \texttt{/youtube/} corpus starting in January 2026:
\begin{enumerate}
  \item \textbf{URL canonicalization.} Duplicate URL patterns (slug-based and video-ID-based)
  serving identical content were consolidated to one canonical URL per video.
  \item \textbf{Demand mining from bot 404 logs.} AI-bot requests to non-existent URLs were treated
  as demand signals; the highest-frequency patterns were used to generate new pages.
  \item \textbf{Title and lead-summary rewriting.} Titles were converted to question form, and lead
  summaries (TLDRs) were rewritten as standalone two-to-three-sentence answers with descriptive
  openers, prioritized by AI-bot request volume per page.
  \item \textbf{SEO-protection rule (``SEO Guard'').} Pages earning meaningful Google clicks in the
  trailing window were locked from rewriting; pages with neither organic nor AI interest were
  unpublished; the remainder entered the rewrite queue.
\end{enumerate}
Because the bundle was applied together, we estimate its \emph{combined} effect, not the marginal
effect of any single tactic. The treatment also changed corpus size (new pages from 404 mining),
so the estimated effect combines per-page improvement and corpus expansion.

\section{Measurement Methodology}
\paragraph{Why an on-domain control.} ChatGPT referral traffic grew across the entire web during
the study window. A before/after comparison on treated pages alone cannot separate treatment from
this tailwind. The untreated remainder of the same domain is exposed to the identical tailwind, so
the \emph{difference} in growth isolates the treatment-attributable component.

\paragraph{Estimator.} Let $R(t) = \text{treated}(t)/\text{control}(t)$. Under the assumption that
platform shocks affect both groups multiplicatively and equally, $\log R(t)$ is invariant to those
shocks. On weekly data we fit the segmented regression
\begin{equation}
\log R_t = \beta_0 + \beta_1 t + \beta_2 D_t + \beta_3 (t-t_0)D_t + \varepsilon_t,
\end{equation}
where $D_t=\mathbf{1}[t \ge t_0]$ and $t_0$ is the first week of January 2026. Here $\beta_1$ is the
pre-existing trend, $\exp(\beta_2)$ is the immediate multiplicative \emph{level} change at the
intervention, and $\beta_3$ is the change in slope. Standard errors are HAC (Newey--West, lag~4).
The pre-period is H2~2025 (the first half of 2025 is excluded because ChatGPT referral volume was
nascent and ratios are dominated by small-count noise).

\paragraph{Inference safeguards.} Because the pre-period is short and autocorrelated, we complement
the regression with (i) a moving-block bootstrap CI for the end-state effect and (ii) a
placebo-in-time permutation test that re-estimates $\beta_2$ at every candidate break inside the
pre-period and asks how unusual the observed January break is.

\section{Results}
\subsection{The raw multiple overstates the effect}
Read trough-to-peak, the daily series rises from its lowest day (early January 2026) to a single-day
peak (5~May~2026) by a factor of $38.2\times$. This is a fragile basis for an effect estimate: the
May~5 value is a transient spike and the sustained late-May level is under half that peak. We
therefore report monthly aggregates instead, on which total ChatGPT referrals grew $5.7\times$, and
treat such raw before/after multiples as descriptive rather than causal.

\subsection{The natural experiment}
\begin{figure}[t]\centering
\includegraphics[width=.8\linewidth]{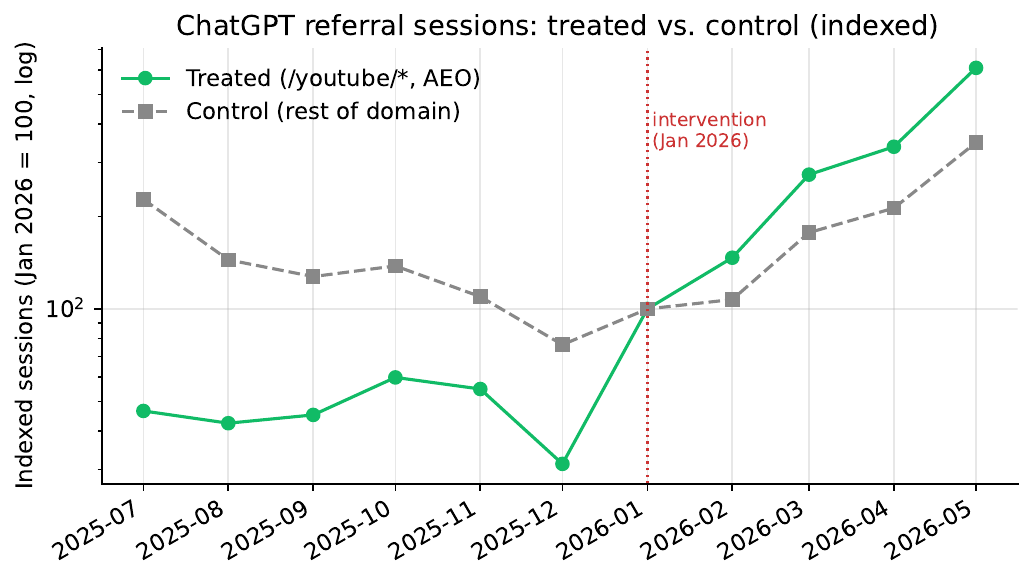}
\caption{ChatGPT referral sessions, treated vs.\ control, indexed to January 2026 $=100$ within
each group (log scale; absolute counts withheld). Both rise, but the treated corpus diverges after
the January 2026 intervention.}
\label{fig:tvc}
\end{figure}

Table~\ref{tab:monthly} and Figure~\ref{fig:tvc} show the split. Within the treatment window,
treated pages grew $6.1\times$ (Jan~$\to$~May 2026) while control pages grew $3.5\times$ from the
same tailwind. The simple difference-in-differences (ratio of growth rates) is $\times 1.75$ on
sessions and $\times 2.48$ on engaged sessions.

\begin{table}[h]\centering
\caption{Monthly ChatGPT referral sessions by group, indexed to January 2026 $=100$ within each
group (absolute counts withheld for confidentiality). $R$ is the treated/control ratio of the
underlying absolute volumes.}
\label{tab:monthly}
\begin{tabular}{lrrr}
\toprule
Month & Treated (index) & Control (index) & Ratio $R$ \\
\midrule
2025-07 & 46.5 & 227.4 & 0.95 \\
2025-10 & 59.8 & 138.0 & 2.02 \\
2025-12 & 31.3 & 76.5 & 1.91 \\
\textbf{2026-01} & \textbf{100.0} & \textbf{100.0} & \textbf{4.67} \\
2026-03 & 273.8 & 177.2 & 7.21 \\
2026-05 & 609.8 & 348.5 & 8.16 \\
\bottomrule
\end{tabular}
\end{table}

\subsection{Interrupted time series}
\begin{figure}[t]\centering
\includegraphics[width=.85\linewidth]{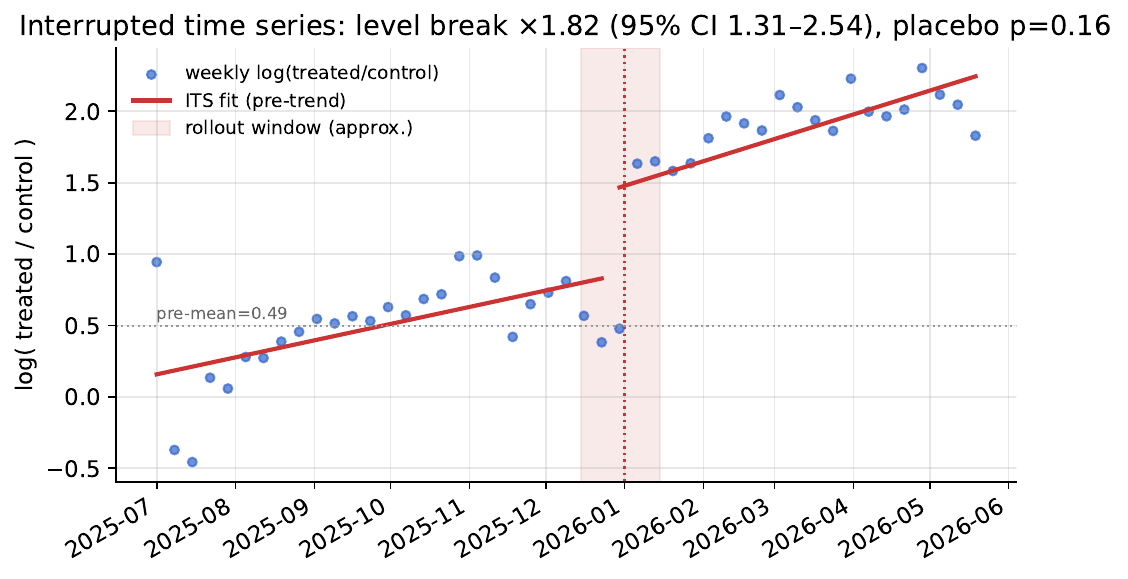}
\caption{Interrupted time series on weekly $\log(\text{treated}/\text{control})$. A significant
pre-trend (red line) continues across a discrete level break of $\times1.82$ at the intervention; the
shaded band marks the approximate rollout window (Dec 2025--Jan 2026), matching the imprecise $t_0$.}
\label{fig:its}
\end{figure}

On the weekly $\log R$ series (47 weeks; 26 pre, 21 post; Figure~\ref{fig:its}):
\begin{itemize}
  \item \textbf{Pre-trend} $\beta_1$: $+0.027$/week ($\times 1.027$, $\approx +11\%$/month),
  $p{=}0.007$. The treated corpus was \emph{already} gaining ChatGPT-referral share before the
  intervention.
  \item \textbf{Level break} $\exp(\beta_2) = \mathbf{1.82}$ (95\% CI $1.31$--$2.54$), $p{=}0.001$:
  a significant discrete jump aligned with the intervention.
  \item \textbf{Slope change} $\beta_3$: $p{=}0.53$ (not significant). The pre-existing trend
  continued; the intervention shifted the level, not the slope.
  \item \textbf{End-state} ratio-of-ratios (last 4 post weeks vs.\ pre-mean): $4.85\times$
  (block-bootstrap 95\% CI $2.63$--$5.55$). This conflates the level break with the continuation
  of the pre-existing trend and should not be read as the intervention effect.
\end{itemize}

\subsection{Placebo test: suggestive, not conclusive}
\begin{figure}[t]\centering
\includegraphics[width=.7\linewidth]{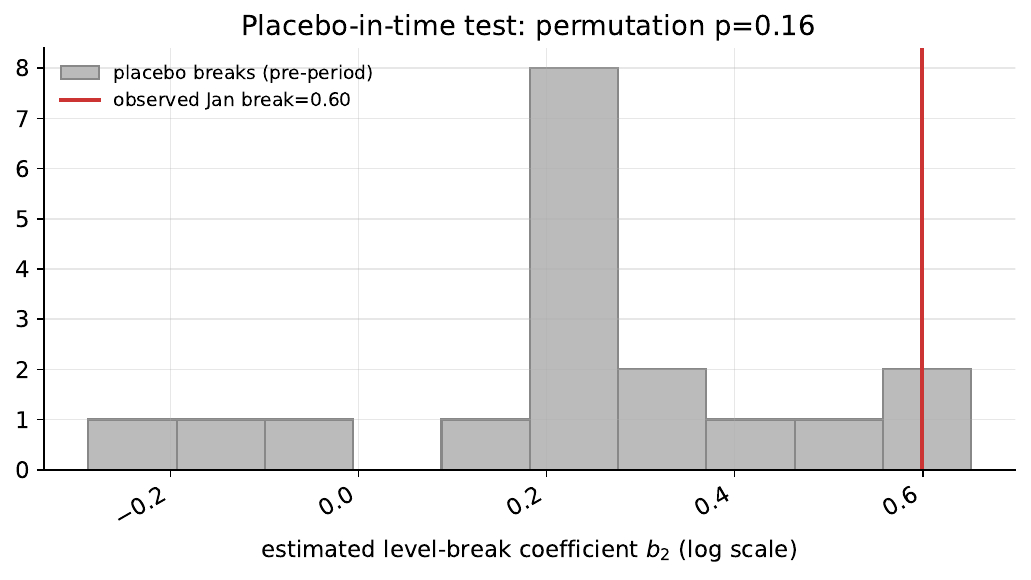}
\caption{Placebo-in-time test. The observed January break (red) is large but not clearly outside the
distribution of spurious pre-period breaks ($p{=}0.16$).}
\label{fig:placebo}
\end{figure}

Re-estimating the level break at every candidate date inside the pre-period yields a placebo
distribution (Figure~\ref{fig:placebo}) with mean $0.24$ and max $|\cdot|=0.65$ against an observed
break of $0.60$. The permutation $p$-value is $\mathbf{0.16}$. The short, volatile pre-period can
produce spurious jumps nearly as large as the observed one, so although the regression effect is
significant under HAC errors, the most conservative test does \emph{not} clear the conventional
threshold. We therefore characterize the effect as suggestive.

\subsection{Robustness}
The level break is stable across specifications: engaged sessions $\times 2.27$ ($p{=}0.001$);
excluding the early-May spike weeks $\times 1.83$ ($p{=}0.001$); and an alternative pre-window
starting 2025-09 $\times 2.07$ ($p{=}0.0002$). All estimates sit in the $1.8$--$2.3\times$ range.

\subsection{SEO safety}
\begin{figure}[t]\centering
\includegraphics[width=.8\linewidth]{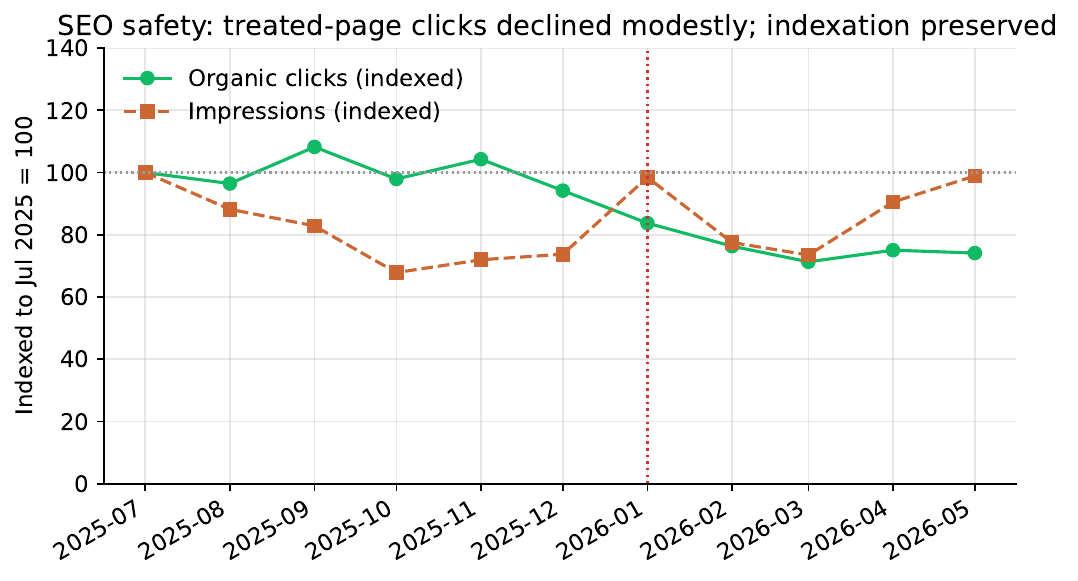}
\caption{SEO safety: treated-page organic clicks and impressions, indexed to July 2025 $=100$
(absolute volumes withheld). Organic clicks declined modestly, in line with the site-wide trend;
impressions remained near baseline (no deindexation).}
\label{fig:seo}
\end{figure}

Google organic clicks to treated \texttt{/youtube/} pages fell only $\approx 25\%$ from their
H2~2025 level to 2026, while impressions stayed within their normal range and ended the window near
baseline (no deindexation; Figure~\ref{fig:seo}). Site-wide organic clicks fell $\approx 20\%$ over
the same window. The treated-page decline is in line with the ambient site-wide and
industry trend rather than a treatment-specific collapse, consistent with the SEO-protection rule.
We find \emph{no} support for the stronger ``flywheel'' claim (that AEO citations measurably lifted
organic search) in this window: treated organic was flat-to-declining, not rising.

\subsection{Engagement composition}
ChatGPT-referral engagement rate rose from $0.486$ (Jan) to $0.896$ (May), with the jump
concentrated at a mid-March bot-filtering policy change. The increase was larger for treated pages
than control, consistent with both better content--query matching and a composition change from
filtering; the ratio/DiD design is intended to be robust to the common component.

\section{Threats to Validity}\label{sec:threats}
\begin{enumerate}
  \item \textbf{Pre-trend present and significant.} $\log R$ drifts upward before treatment.
  Mitigation: the ITS explicitly models the pre-trend and the placebo test bounds spurious breaks;
  we report the placebo $p{=}0.16$ honestly rather than relying on the HAC $p$ alone.
  \item \textbf{SUTVA / spillover.} If AEO raised domain-level citation propensity, untreated pages
  also benefited, inflating the control and biasing DiD \emph{downward}; reported effects are then
  conservative.
  \item \textbf{Measurement discontinuity.} The mid-March bot-filtering change altered session
  composition. Mitigation: engaged-session outcome and common-shock differencing.
  \item \textbf{Bundle, not component.} The four tactics are confounded with one another.
  \item \textbf{Single domain, single engine.} One site, ChatGPT-dominant; external validity
  untested.
  \item \textbf{Content heterogeneity between groups.} Treated and control differ in intent and are
  not matched.
  \item \textbf{Observational, non-randomized; imprecise rollout date} ($t_0$ approximated as
  January 2026).
\end{enumerate}

\section{Discussion}
The practical lesson is that \textbf{the platform tailwind is large and must be netted out.} On this
domain, untreated pages grew $3.5\times$ with no optimization at all; a naive before/after on
treated pages would credit the full $6.1\times$ to the optimization. The defensible,
intervention-aligned level effect is $\sim$1.8--2.3$\times$,
and even that is only suggestive under the most conservative test. We suspect many published AEO
multiples are similarly inflated by conflating treatment with tailwind, and that on-domain controls
are the lowest-cost available correction. A secondary lesson is that AEO and SEO need not be in
tension: a guarded rewrite program improved answer-engine referrals without measurable organic harm.

\section{Limitations and Future Work}
A randomized second-pass rewrite (holding out a random subset of eligible pages) would convert the
structural-break argument into a clean experimental effect; we did not run one here. Matched
controls, multi-engine and multi-domain replication, and an external platform-wide referral index
would each strengthen identification.

\section{Conclusion}
Using an on-domain control and first-party logs, we separate AEO treatment effect from platform
growth on a large corpus (hundreds of thousands of pages). The credible, platform-adjusted level
effect is $\sim$1.8--2.3$\times$ (suggestive, $p_{\text{placebo}}
{=}0.16$), achieved without measurable damage to organic search, against a $3.5\times$ platform
tailwind that dominates the raw numbers. The contribution is a low-cost, reproducible measurement
design that we believe should be standard when reporting AEO results.

\appendix
\section{Reproducibility}
The monthly and weekly series, the analysis script (OLS with a hand-implemented Newey--West
estimator, the ITS model, the placebo test, and robustness checks), and the plotting script are
provided as supplementary material (\texttt{data/}, \texttt{stats.py}, \texttt{plot.py}). Data were
queried from the GA4 Data API and the Search Console API on 2026-05-29.

\end{document}